\documentclass[aps,prb,twocolumn,superscriptaddress,amsmath,amssymb,groupedaddress]{revtex4-2}
\usepackage{ragged2e}
\usepackage{caption}
\DeclareCaptionJustification{justified}{\justifying}
\usepackage[colorlinks,allcolors=cyan]{hyperref}
 
\usepackage{graphicx}
\usepackage{dcolumn}
\usepackage{bm}
\usepackage[capitalise]{cleveref}
\usepackage{physics}
\usepackage{xcolor}

\usepackage{comment}
\usepackage{bbold}
\usepackage{mathtools}
\usepackage{bbold}
\usepackage[caption=false]{subfig}
\DeclareCaptionOption{justified}[]{\caption@setjustification{justified}}

\def\br{{\bf r}}

\newif\ifSM
\SMtrue
\newif\ifmainText
\mainTexttrue 

\begin{document}
\def\lc{\left\lfloor}   
\def\rc{\right\rfloor}
\setlength{\intextsep}{10pt plus 2pt minus 2pt}

\ifmainText
\title{Trimer superfluidity of antiparallel dipolar excitons in a bilayer heterostructure}

\author{
Pradyumna P. Belgaonkar$^{1}$, Michal Zimmerman$^{1}$, Snir Gazit$^{1,2}$, and Dror Orgad$^{1}$
} 

\affiliation{
$^{1}$Racah Institute of Physics, The Hebrew University of Jerusalem, Jerusalem 9190401, Israel \\
$^{2}$The Fritz Haber Research Center for Molecular Dynamics, The Hebrew University of Jerusalem, Jerusalem 9190401, Israel
}

\date{\today}

\begin{abstract}
We study the phase diagram of a bilayer of antiparallel dipolar excitons with a 1:2 density ratio between the layers, 
as a function of temperature and density. Using quantum Monte Carlo simulations, we show that such a system supports
the formation of trimers, namely, three-exciton bound states consisting of a single dipole in one layer and two dipoles
in the second layer. At sufficiently low temperatures and densities, these trimers condense into a trimer superfluid phase. 
Increasing the excitonic density induces a quantum phase transition into a phase in which condensates of independent dipoles 
exist in both layers, in parallel to the trimers. We also study the thermal transitions out of these phases, 
and find that while the normal state is reached directly from the trimer superfluid, the thermal disordering of the 
two-superfluid phase involves an intermediate state which is either a trimer superfluid or a single excitonic condensate 
in the denser layer. A potential experimental realization using 
transition metal dichalcogenide heterostructures is discussed.

\end{abstract}
\maketitle

\section{Introduction}
Dipolar interactions in quantum many-body systems give rise to a rich variety
of exotic collective behavior. Their anisotropic character, where forces can be either repulsive or 
attractive depending on the relative dipolar orientation and inter-particle separation, coupled with 
their long-range nature, can stabilize unconventional excitations and promote strong correlations.

Dipolar systems have been extensively investigated experimentally in ultracold atomic gases with either 
magnetic or electric dipole moments  \cite{Lahaye_2009,Chomaz_2022}. In the low temperature regime, experiments 
have demonstrated Bose-condensed states \cite{Tilman_2005,Lev_2011,Ferlaino_2012} and, more recently, 
supersolid phases \cite{Ferlaino_2019,Modugno_2019,Tilman_2019}. In the solid-state realm, dipolar excitons constitute 
a particularly promising platform for enhancing correlations due to their relatively high density and pronounced 
electrostatic interactions. Achieving a permanent dipole moment requires spatial separation of electrons and holes. 
In semiconductor heterostructures, this is often achieved in a double quantum well subjected to an external 
electric field, which orients the dipoles \cite{Butov_2004,Rapaport_2007}. However, an emerging alternative 
utilizes transition metal dichalcogenide (TMD) heterostructures, where the layering itself determines the 
dipole orientation, even in the absence of an aligning field \cite{Paik_24}.

Exposing the anisotropic nature of dipolar interactions requires structures that go beyond a single layer of dipoles. 
A simple extension involves stacking a pair of double quantum wells, as realized in a semiconducting quadrilayer 
systems \cite{Rapaport_2019,Gossard_2021}. Indeed, the attractive interlayer interaction between parallel dipoles 
was observed experimentally. Interestingly, a sufficiently strong attraction is predicted to support the formation 
of a dimer superfluid comprising bound interlayer molecules \cite{Zimmerman_2022}.

The versatility of TMD-based systems opens up the possibility to engineer the dipole orientation of each 
layer independently. In particular, arranging antiparallel dipoles results in inverted interlayer interactions 
relative to the previously studied parallel case. The collective behavior of quantum antiparallel dipoles holds 
the promise of realizing new phases of composite particles.

\begin{figure}[t]
    \centering
    \includegraphics[width=1.0\linewidth]{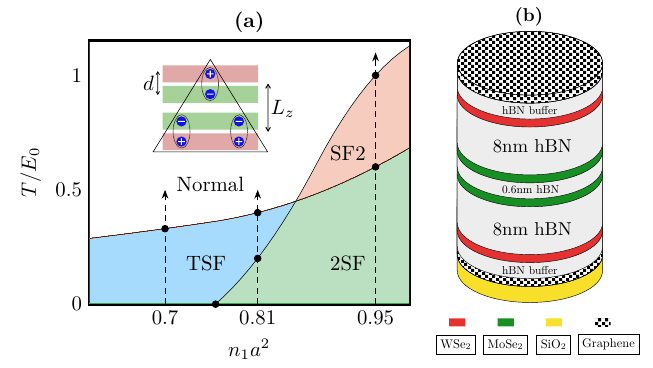}
    \captionsetup[subfigure]{labelformat=empty}
    \subfloat[\label{subfig:phase_diagram}]{}
    \subfloat[\label{subfig:TMD}]{}
    \caption{(a) The phase diagram of a bilayer of antiparallel dipolar excitons with 1:2 density ratio, as depicted 
    in the inset. The phases include a trimer superfluid (TSF), independent superfluids in the two layers (2SF) and 
    a superfuid in layer 2 (SF2). The critical temperatures along the dashed lines were calculated for the TMD-based 
    heterostructure whose parameters are specified in (b). The graphene layers enable electrical gating.}
    \label{fig:Model}
\end{figure}

In this work, we investigate a quadrilayer structure that realizes a bilayer of oppositely oriented dipoles. 
Using numerically exact large-scale quantum Monte Carlo simulations, we determine the low-temperature phase 
diagram of such a system, specifically at a 1:2 filling ratio between the layers. In the vicinity of this point, 
excitons tend to form trimers, namely three-exciton interlayer bound states that exploit the attractive component 
of the dipolar interaction. At low temperatures and in the dilute limit, these trimers condense, forming 
a phase that partially breaks the global $U(1) \times U(1)$ symmetry down to $U(1)$. Upon increasing the 
density, we identify a zero-temperature BEC quantum phase transition, where the residual $U(1)$ symmetry 
breaks. The finite-temperature phase diagram exhibits a rich structure featuring 
Berezinskii–Kosterlitz–Thouless (BKT) transitions between composite and elementary superfluids. 
In addition, we also comment on the nature of the phase diagram away from the 1:2 point, discuss 
the experimental aspects involved in realizing the model in TMD heterostructurs, and outline observable 
signatures of trimers and their condensation.

\section{Model}
We model the many-exciton system within the quadrilayer structure shown in \cref{fig:Model}
as a collection of point-like dipolar bosons arranged in two layers separated by distance $L_z$. 
This approximation is valid in the dilute limit, where the extent of the exciton wavefunction is small 
compared to the inter-exciton separation. The dipole moments in the top (bottom) layer are pointed in the 
positive (negative) $\hat{z}$ direction, due to the mirror symmetry of the quadrilayer. 
Neglecting tunneling between layers, the dynamics follows the Hamiltonian

\begin{equation}
\begin{aligned}
    \mathcal{H}&=-\frac{\hbar^2}{2m_X}\sum_{i,\alpha} \nabla^2_{i,\alpha}+\frac{D^2}{2}
    \sum_{i_\alpha\ne j_\beta} V\qty(\vb{r}_{i_\alpha},\vb{r}_{j_\beta}),
\end{aligned}
\label{eq:H}
\end{equation}
where $m_X$ is the effective exciton mass and the dipole strength $D^2=e^2 d^2 / \varepsilon$ 
is given in terms of the exciton dipole moment $ed$ and the permittivity $\varepsilon$. 
We label excitons by $i_\alpha$, with $\alpha=1,2$ denoting their layer index, and $\vb{r}_{i_\alpha}$ their 
{\it in-plane} position. Assuming that the excitonic densities $n_\alpha=N_\alpha/A$ obey $n_\alpha d^2\ll 1$ 
their dipolar interaction takes the form 
\begin{equation}\label{eqn:V_dipole}
V\qty(\vb{r}_{i_\alpha},\vb{r}_{j_\beta}) = \begin{cases}
1/\qty|\vb{r}_{i_\alpha}-\vb{r}_{i_\beta}|^{3}, & \alpha=\beta\\
-\frac{\qty|\vb{r}_{i_\alpha}-\vb{r}_{i_\beta}|^{2}-2L_z^{2}}{\left(\qty|\vb{r}_{i_\alpha}-\vb{r}_{i_\beta}|^{2} 
+ L_z^{2}\right)^{\frac{5}{2}}}, & \alpha\ne\beta.
\end{cases}
\end{equation}
Henceforth, we measure lengths and energies in units of $a=m_X D^2/\hbar^2$ and $E_0=D^2/a^3$, respectively. 
To maintain contact with the physical systems, we note that $a=183$ nm and $E_0=4.54$ $\mu$eV for a $d=8$ nm 
MoSe$_2$-WSe$_2$ structure \cite{Rivera2015,Rivera2016,Jauregui2019,Zhang2019,Liu2021,Brotons2021}, see \cref{subfig:TMD}.

\section{Three-body bound state}
We begin our analysis by establishing the presence of a trimer bound state.
Crucial for its formation is the spatial structure of the interlayer interaction. Initially repulsive 
at short distances, it becomes attractive for separations larger than $\sqrt{2}L_z$. This allows the two 
excitons in the bottom layer to bind to the exciton in the top layer while avoiding their strong short-distance 
intralayer repulsion. The total interaction potential seen by the bottom layer excitons attains a minimum when 
their mutual separation is $4.28L_z$. This distance may be taken as a rough estimate for the diameter of the trimer. 
It is useful to contrast the above discussion with the case of parallel dipoles \cite{Zimmerman_2022}, where 
the interlayer attraction peaks at zero separation, precisely where the intralayer repulsion diverges.

To substantiate the above scenario, we diagonalized numerically the Hamiltonian, Eq. (\ref{eq:H}), 
and obtained the three-exciton ground state. The results, shown in Fig. \ref{fig:3body}, clearly demonstrate 
a trimer bound state, whose binding energy decreases monotonically with $L_z/a$. We present its 
structure using the coordinates $\br_+=(\br_2+\br_3)/2-\br_1$ and $\br_-=(\br_2-\br_3)/2$, defined in terms 
of the positions $\br_1$ and $\br_{2,3}$ of the excitons in the upper and lower layers, respectively. 
The probability distribution function $|\psi_+(\br_+)|^2=\int d^2r_-|\psi(\br_+,\br_-)|^2$ is centered 
at $\br_+=0$, indicating a symmetric configuration of the lower particles around $\br_1$, see Fig. \ref{subfig:psi_minus}. 
Concomitantly, $|\psi_-(\br_-)|^2=\int d^2r_+|\psi(\br_+,\br_-)|^2$ reveals that $\br_{2,3}$ largely occupy antipodal 
points on a ring, whose diameter for the case $L_z/a=0.047$ shown in Fig. \ref{subfig:psi_plus} is 60\% 
larger than the classical trimer diameter. This is due to the slow decay of the interaction potential 
at large separations.

\begin{figure}[t]
    \centering    \includegraphics[width=1.0\linewidth]{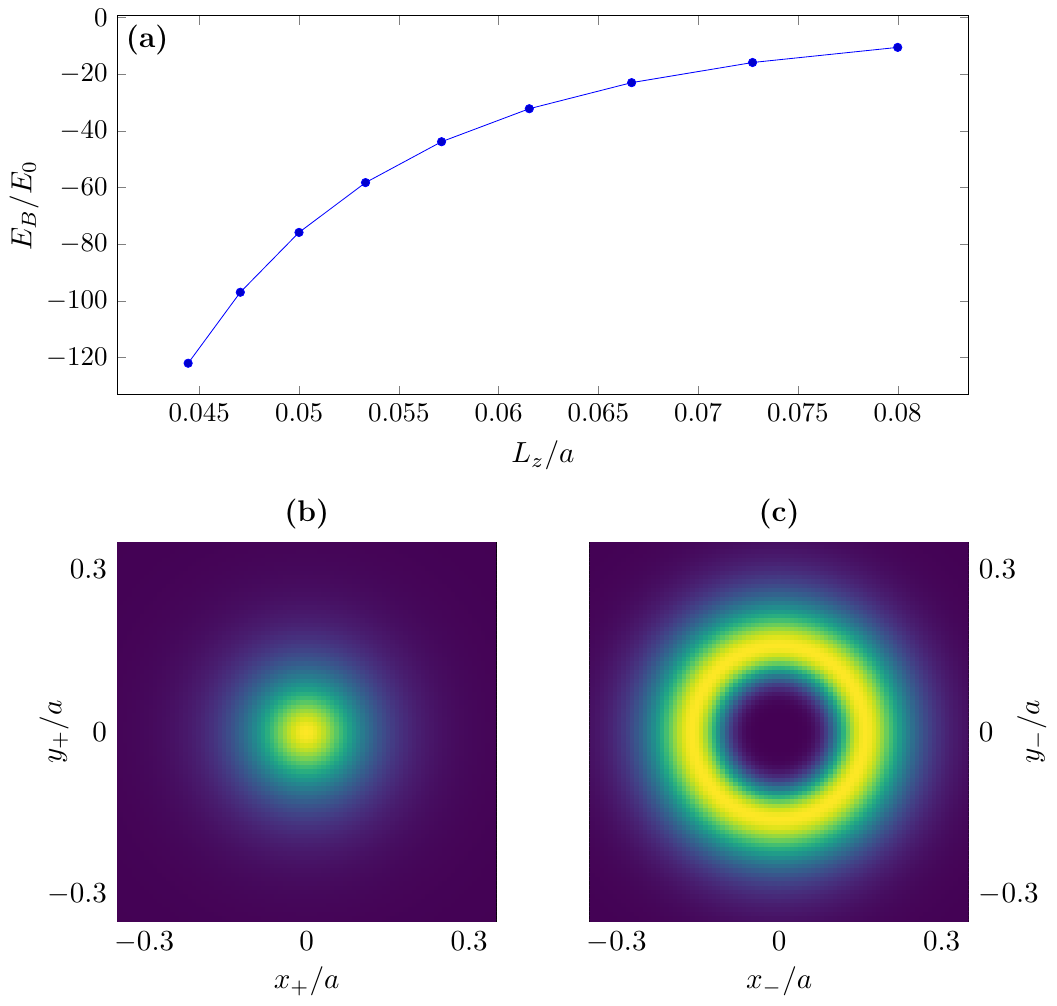}
    \captionsetup[subfigure]{labelformat=empty}
    \subfloat[\label{subfig:binding_energy}]{}
    \subfloat[\label{subfig:psi_minus}]{}
    \subfloat[\label{subfig:psi_plus}]{}
    \caption{The trimer bound state. (a) The binding energy as a function of $L_z/a$. (b) and (c) 
    The probability distribution functions $|\psi_+(\br_+)^2|$ and $|\psi_-(\br_-)|^2$ for the case $L_z/a=0.047$.}
    \label{fig:3body}
\end{figure}

The interlayer interaction enables the formation of other bound states comprising one upper-layer exciton 
and $n-1$ lower-layer excitons. Classically, the minimum energy for such an $n$-mer is achieved in a static 
configuration where the upper exciton is positioned directly above the center of a regular $(n-1)$-sided polygon, 
with the lower-layer excitons at its vertices.
The binding energy is expected to diminish with $n$ because the number of repulsive 
lower-layer pairs grows quadratically as $\binom{n-1}{2}$, whereas the attractive interlayer component scales 
only as $n-1$. Indeed, we find that the classical binding energy maximizes for the trimer and subsequently 
decreases for $n\geq 4$. The classical binding-energy of the dimer is approximately $40\%$ smaller than that of the trimer, 
a consequence of the trimer's additional attractive interlayer interaction overcoming its intralayer repulsion 
for sufficiently separated lower-layer excitons. Quantum mechanically, the uncertainty principle smears the classical 
configuration, introducing a kinetic energy contribution that grows with $n$ and reduces the binding energy. 
Consequently, the quantum binding energy of the dimer is only $20\%$ smaller than that of the trimer 
(see Appendix \ref{app:boundstate}), but the trimer still emerges as the most favorable complex of 
anti-parallel excitons within the bilayer structure.

\begin{figure*}[t!]
    \centering
    \includegraphics[width=0.96\linewidth]{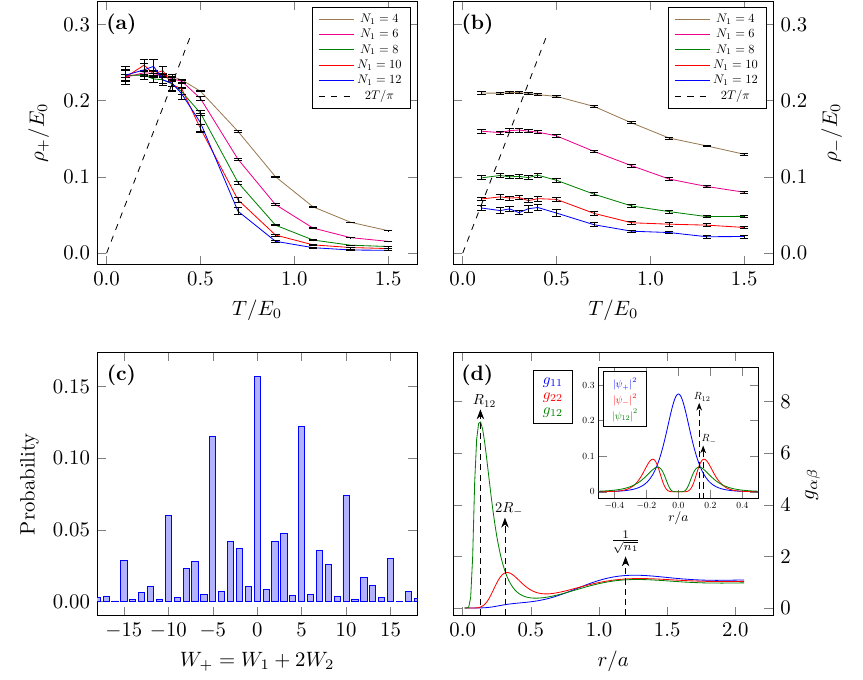}
    \captionsetup[subfigure]{labelformat=empty}
    \subfloat[\label{subfig:rhoplus}]{}
    \subfloat[\label{subfig:rhomin}]{}
    \subfloat[\label{subfig:Wplus}]{}
    \subfloat[\label{subfig:dencorr}]{}
    \caption{Evidence for a TSF in a system with $L_z/a=0.047$ and $n_1=0.7/a^2$. (a) and (b) $\rho_+$ 
    and $\rho_-$ as a function of temperature for various numbers, $N_1$, of excitons in layer 1. 
    (c) The distribution of the winding number $W_+$ for the case  $N_1=12$ and $T=0.1E_0$.
    (d) The intralayer ($g_{11}$, $g_{22}$) and the interlayer ($g_{12}$) density-density
    correlation functions for the same case. They attain a maximum at $1/\sqrt{n_1}$, $2R_-$ and $R_{12}$, 
    respectively. $R_-$ and $R_{12}$ are the positions of the maxima of $|\psi_-|^2$ and $|\psi_{12}|^2$, 
    as shown in the inset.}
    \label{fig:TSFevidence}
\end{figure*}

\section{Trimer condensate and its density driven quantum phase transition} 
In light of the above discussion, a simple energetic argument 
suggests the existence of a trimer liquid at the commensurate ratio $n_2/n_1=2$ between the excitonic 
densities, at least in the dilute limit $n_\alpha a^2\ll 1$ where the interaction between trimers is weak and 
is likely unable to destabilize the bound states. Consequently, we focus on the case $n_2/n_1=2$, and comment 
on the physics away from this point in the Discussion section. 

At sufficiently low temperatures below a BKT transition, the trimers are 
expected to condense into a trimer superfluid state (TSF). At zero temperature the TSF is characterized by 
partial breaking of the global $U(1)\times U(1)$ symmetry, associated with the independent conservation of 
particle number in each layer. Specifically, the TSF order parameter 
$\mathcal{O}_{\text{TSF}}= \expval{b_1 b_2 b_2}=\sqrt{\rho_T} e^{i\theta_+}$, expressed in terms of the exciton 
annihilation operators $b_\alpha\sim e^{i\theta_\alpha}$, acquires a nonzero vacuum expectation value, which 
locks the phase variable $\theta_+=\theta_1+2\theta_2$. By contrast, the complementary angle variable, 
$\theta_-=2\theta_1-\theta_2$, fluctuates and together with  $\theta_{1,2}$, remains disordered. 

In the dense limit $n_\alpha a^2\gg 1$ the repulsive intralayer interaction dominates over the interlayer 
attraction and is expected to cause dissociation of the trimers. In this limit the layers decouple 
and can condense into two independent superfluids, thus breaking the remaining $U(1)$ symmetry
and establishing order in both $\theta_+$ and $\theta_-$ (or equivalently in $\theta_{1,2}$.) We note, however, 
that the transition out of the TSF at $n_1=n_c$ does not necessarily imply significant dissociation of 
the bound states. 
Instead, independent superflow of excitons may emerge in parallel to the superflow of trimers,  
without compromising the integrity of the latter.
Such a phase exhibits the same 
symmetry breaking pattern as the state of two completely independent superfluids, and we denote them both by 2SF. 
They can be distinguished by the presence of non vanishing interlayer drag.

\begin{figure*}[t]
    \centering
    \includegraphics[width=0.98\linewidth]{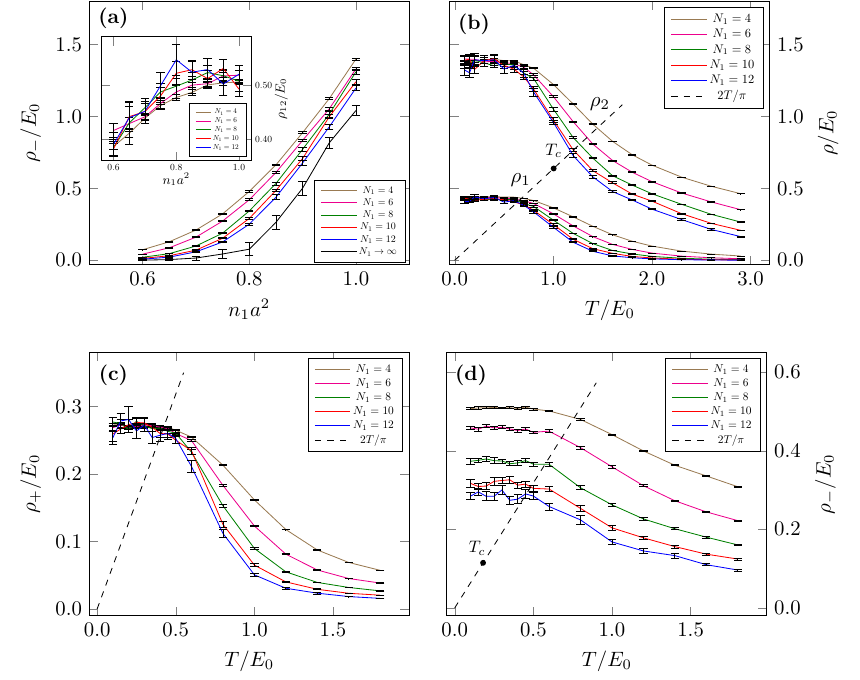}
    \captionsetup[subfigure]{labelformat=empty}
    \subfloat[\label{subfig:FSS}]{}
    \subfloat[\label{subfig:highden}]{}
    \subfloat[\label{subfig:interrhoplus}]{}
    \subfloat[\label{subfig:interrhominus}]{}
    \caption{Finite size scaling of $\rho_-$ as a function of $n_1$ for a system with $L_z/a=0.047$ 
    at $T=0.1E_0$. The inset depicts the cross-stiffness $\rho_{12}$. (b) $\rho_{1,2}$ as a function
    of temperature for a system with $n_1=0.95/a^2$. (c) and (d) $\rho_+$ and $\rho_-$ as a function 
    of temperature for a system with $n_1=0.81/a^2$. The critical temperatures indicated in (b) and (d) 
    are obtained by finite size scaling of the data for $\rho_2$ and $\rho_-$, respectively, as shown 
    in Appendix \ref{app:Tcscale}.}
    \label{fig:QPT}
\end{figure*}

To elucidate the properties of the many-exciton system we use path integral Monte Carlo (PIMC) simulations 
\cite{Ceperley_1995}, and specifically the worm algorithm \cite{Boninsegni_2006}. We establish the ground 
state properties by tracking the convergence of the finite-temperature results when we lower the  
temperature $T$ and increase the total number of particles to $N_{\text{tot}}=36$. To detect the onset 
of order 
we calculate the superfluid stiffness 
tensor $\rho_{\alpha\beta}=\partial^2 F/\partial\Phi_\alpha\partial\Phi_\beta$, measuring the response 
of the free energy $F$ of a system on a torus to the insertion of Aharonov-Bohm fluxes,  
$\Phi_\alpha$, minimally coupled to layer $\alpha$. An eigenmode of $\rho_{\alpha\beta}$ with a
nonzero eigenvalue in the $T=0$, $N\rightarrow \infty$ limit, corresponds to the phase of an order parameter that breaks a $U(1)$ symmetry, 
while a zero-mode is associated with a conserved $U(1)$ charge.  In practice, we calculate the 
stiffness via the winding numbers ${\bf W}_\alpha=\{W_\alpha^x,W_\alpha^y\}$ of the world lines of 
layer-$\alpha$ particles around the $x,y$ directions using 
$\rho_{\alpha\beta}=T\langle {\bf W}_\alpha \cdot{\bf W}_\beta\rangle/2$ \cite{Pollock_1987}. 
The phases $\theta_\pm$ couple to $\Phi_+=\Phi_1+2\Phi_2$
and $\Phi_-=2\Phi_1-\Phi_2$. Correspondingly, the stiffness of $\theta_+$ is given by 
$\rho_+=T\langle ({\bf W}_1+2{\bf W}_2)^2\rangle/50$. Whenever $\theta_+$ orders, the vortices responsible  
for disordering $\theta_-$ involve $10\pi$ phase windings (see next section and Appendix \ref{app:xy}). 
The stiffness that is relevant to their proliferation is given by the response to $\Phi_-/5$ and takes the 
form $\rho_-=T\langle (2{\bf W}_1-{\bf W}_2)^2\rangle/2$.

Fig. \ref{fig:TSFevidence} provides evidence for a TSF phase. In particular, it depicts 
the temperature dependence of $\rho_\pm$ for a system with $L_z/a=0.047$ and $n_1=0.7/a^2$. Clearly, 
the low-temperature $\rho_+$ remains above the BKT line $\rho=2T/\pi$ as the number of excitons increases, 
signaling the establishment of $\theta_+$ quasi-long-range order (QLRO). Conversely, the finite size 
scaling of $\rho_-$ indicates  its disappearance in the thermodynamic limit, see Appendix \ref{app:datacoll}. 
Further supporting evidence comes from the distribution of $W_+=W_1+2W_2$ that peaks at multiples of 5, 
indicating the correlated winding of excitons as composites comprising a single exciton in the 
top layer ($W_1=1$) and a pair of excitons in the bottom layer ($W_2=2$).

Finally, we evaluate the density-density correlation functions 
$g_{\alpha\beta}(\br)=A/(N_\alpha N_\beta) \sum_{i_\alpha\neq j_\beta}\langle\delta(\br-\vb{r}_{i_\alpha}-\vb{r}_{j_\beta})\rangle$.
As expected from a liquid of trimers, $g_{22}$ attains a maximum at the trimer diameter, which is 
identified from the peak in $|\psi_-|^2$, while $g_{12}$ reaches its maximum at the same position as 
the probability distribution $|\psi_{12}(\br_1-\br_2)|^2=\int d^2r_3|\psi(\br_1-\br_2,\br_1-\br_3)|^2$.

In Fig. \ref{subfig:FSS} we present the finite size scaling of the low-temperature $\rho_-$ as a function 
of density. This, together with scaling of the superfluid fraction in the two layers shown in 
Appendix \ref{app:sffrac}, enable us to identify a quantum phase transition (QPT) from the 
TSF into a state where $\theta_-$ orders as well. For the system we have studied numerically 
the transition occurs at a critical density $n_1=n_c\approx 0.76/a^2$. 

Figure \ref{subfig:FSS} also shows that the cross-layer stiffness $\rho_{12}$, which is responsible 
for superfluid drag \cite{Babaev}, initially increases with density and then levels off for $n_1>n_c$. 
This observation, alongside the results for $g_{\alpha\beta}(\br)$ presented in Appendix \ref{app:dencorr}, 
supports the scenario where trimers largely maintain their integrity beyond the QPT. For $n_1>n_c$, 
parallel independent superfluidity in both layers is established via hops of single excitons between 
trimers, accompanied by a global rearrangement of the other particles into new trimers 
(see movies \cite{zenodo} and their discussion in Appendix \ref{app:windings}). 
The persistence of trimers in the 2SF phase and the resulting large dissipationless drag stand  
in contrast to the behavior of a bilayer of parallel excitons with equal densities, where dimers 
dissociate and $\rho_{12}$ rapidly diminishes in the 2SF phase \cite{Macia}. Furthermore, the  
exchange symmetry between the layers endows the equal-density transition with a 
relativistic ($z=1$) character \cite{Zimmerman_2022}. In contrast, the system considered here 
is expected to undergo a more conventional $z=2$ superfluid transition. The scaling analysis of 
$\rho_-$, included in Appendix \ref{app:datacoll}, gives evidence to this effect.


\section{Finite temperature phase diagram}
Having addressed the zero temperature behavior, we now turn to examine
the system's thermal phase transitions. The picture for $n_1<n_c$ is simple, with a BKT disordering transition of 
$\theta_+$ from the TSF to a normal $U(1)\times U(1)$ symmetric state, as shown by Fig. \ref{subfig:rhoplus}. 
To identify possible scenarios for $n_1>n_c$ let us consider independent condensates in each layer as well as 
a condensate of trimers. The thermal fluctuations of their phases, $\theta_{1,2}$ and $\theta_T$, are governed by 
an XY coupling $\sum_\mu\cos(\nabla_\mu\theta)$. Additionally, one needs to account for processes converting a 
trimer into unbound particles, represented by a term proportional to $\cos(\theta_T-\theta_1-2\theta_2)$ \cite{Kuklov_2004}, 
which also reduces the symmetry to $U(1)\times U(1)$.
At temperatures well below the trimer binding energy and in the vicinity of $n_c$ we replace it 
by a rigid constraint, leading to the following effective Hamiltonian density in terms of $\theta_{1,2}$ only
\begin{eqnarray}
    \label{eq:effXY}
    \nonumber
    {\cal H}&=&-\sum_\mu (n_1-n_c)[\cos(\nabla_\mu\theta_1) + 2\cos(\nabla_\mu\theta_2)] \\
    &&-\sum_\mu\frac{n_c} {3}\cos[\nabla_\mu(\theta_1+2\theta_2)].
\end{eqnarray}
In the above model, $\rho_{12}$ is proportional to the coefficient of the last term. Based on the observation that 
at low temperatures $\rho_{12}$ remains approximately constant for $n_1>n_c$, we have fixed the bare stiffness of 
the trimers in Eq. (\ref{eq:effXY}) to their density to mass ratio at $n_c$ (setting the exciton mass to 1.) 
Concomitantly, the bare stiffness of the excitonic condensates reflects the excess density in the two layers 
above $n_c$.

For $n_1$ close to $n_c$, the last term in Eq. (\ref{eq:effXY}) is expected to dominate and lead at low temperatures 
to QLRO in $\theta_+$. Expressing $\theta_{1,2}$ in terms of $\theta_\pm$ and approximating the QLRO by true 
long-range order in $\theta_+$ we obtain ${\cal H}\propto \sum_\mu[\cos(\nabla_\mu2\theta_-/5)
+2\cos(\nabla_\mu\theta_-/5)]$. The disordering transition of such a Hamiltonian is driven by the proliferation of either 
$5\pi$  or $10\pi$ vortices in $\theta_-$, depending on the coupling constants of the two terms 
\cite{Lee_1985,Carpenter_1989}. In Appendix \ref{app:xy} we present Monte Carlo results showing   
that for small $n_1-n_c$, model (\ref{eq:effXY}) undergoes a BKT transition induced by proliferation of $10\pi$ 
vortices in $\theta_-$. Fig. \ref{subfig:interrhominus} demonstrates that we find a similar 
transition from a 2SF to a TSF also in the PIMC data of model (\ref{eq:H}) with $n_1=0.81/a^2$. This is followed 
at a higher temperature by a transition into the normal state , where the QLRO in $\theta_+$ 
is lost, as seen in Fig. \ref{subfig:interrhoplus}.

With increasing $n_1$, the $\cos(\nabla_\mu\theta_2)$ term in Eq. (\ref{eq:effXY}) gains importance and eventually 
becomes dominant, causing $\theta_2$ to develop QLRO. Neglecting the fluctuations in $\theta_2$ one finds 
${\cal H}\propto \sum_\mu\cos(\nabla_\mu\theta_1)$. Now, the transition out of the low-temperature 2SF phase 
is best characterized in the 1,2 basis, as it involves loosing superfluidity in layer 1. 
The higher temperature phase, denoted as SF2, maintains superfluidity in layer 2 while $\rho_1=\rho_{12}=0$. 
It eventually turns into the normal state where $\rho_2=0$ as well 
($\rho_\alpha\equiv\rho_{\alpha\alpha}$.) Fig. \ref{subfig:highden} illustrates that this scenario materializes 
in the excitonic model with $n_1=0.95/a^2$. Taken together, the above discussion implies the finite-temperature 
phase diagram depicted in Fig. \ref{subfig:phase_diagram}.

\section{Experimental considerations}
As a reference, we consider a MoSe$_2$-WSe$_2$ system
\cite{Rivera2015,Rivera2016,Jauregui2019,Zhang2019,Liu2021,Brotons2021} with $d=8$ nm and $L_z=8.6$ nm, 
see Fig. \ref{subfig:TMD}. We obtain a trimer binding energy $E_B = 0.44$ meV and an average trimer 
radius of 30 nm. We note that a calculation for a MoS$_2$-WS$_2$ bilayer with $d=8$ nm predicts 
an interlayer exciton with an approximate binding energy of 75 meV and a radius of 6 nm \cite{Peeters}.
The significant size difference between the trimer and its constituent excitons justifies treating 
the latter as point particles at the relevant length scales. This statement holds true also since 
our calculation for the same system yields a QPT at $n_c = 2.27 \times 10^9$ cm$^{-2}$, 
and $T_c = 0.017$ K for the TSF-Normal transition at $n_1 = 2.1 \times 10^9$ cm$^{-2}$, corresponding to 
an inter-trimer distance of about 210 nm. 

We find that for fixed $L_z/d$, $E_B$ changes little with $d$ in the range $5-10$ nm, see Appendix 
\ref{app:boundstate}. Within this range, the location of the TSF-2SF QPT is governed by the size of the 
trimers and not by their binding energy, leading to $n_c\sim 1/L_z^2$. Therefore, we can estimate for a 
system with $d=5$ nm and $L_z=6$ nm that $E_B\simeq 0.3$ meV and $n_c\simeq 4\times10^9$ cm$^{-2}$. 
For fixed $d$, both $E_B$ and $n_c$ are increasing functions of $d/L_z$. Thus, in order to maximize $E_B$ 
and $n_c$ (and with it $T_c$ of the TSF-Normal transition) it is preferable to make $L_z$ as small as possible. 

Interlayer excitons in TMD heterostructures are generated via optical excitation of intralayer excitons, 
followed by charge transfer between layers due to their inherent (type-II) band alignment \cite{Paik_24}. 
The significant spatial separation between the electron and hole within these interlayer excitons grants 
them very long lifetimes, on the order of 10 ns \cite{Calman2018}, much longer than the typical 
lifetime of intralayer excitons, which is of the order of 10 ps \cite{Wang-review}. Consequently, 
one can isolate the physics associated with the interlayer excitons by introducing a time delay 
between the optical excitation and the subsequent detection measurement. Furthermore, direct and 
interlayer excitons are energetically separated by hundreds of meV, making them readily distinguishable 
in the light emission spectrum \cite{Calman2018}.

The spatial separation granting interlayer excitons their long lifetimes can also impede their 
recombination and their subsequent optical detection due to the requirement for charge tunneling. Nonetheless,  
such excitons have been observed in a TMD heterostructure with a 1 nm hBN spacer \cite{Calman2018}. In this case, 
direct tunneling would be prohibitively slow, suggesting that tunneling instead proceeds via intermediate defect 
levels within the hBN spacer. Therefore, we propose facilitating tunneling at even larger separations by 
intentionally doping the hBN spacers with impurities \cite{hBN-impurities}. This approach is supported by 
a recent demonstration of defect-assisted tunneling through a 4 nm thick carbon-doped hBN barrier \cite{Seo2024}. 
Another option is to form a superlattice of hBN and TMD layers, where the latter act as a stack of 
effective "defect" states through which tunneling may progress. 

Attaining the desired excitonic-density ratio between the bilayers requires selective excitation, which necessitates 
breaking the symmetry between them. This can be achieved by applying an electric field via top 
and bottom gates. The field modifies the band alignment in the two bilayers and was shown to tune both the energy 
and intensity of photoluminescence from interlayer excitons in TMD bilayers, whether in direct contact 
\cite{Rivera2015} or separated by an hBN spacer \cite{electric-tune}. Another approach, which could 
also be integrated with an applied field, involves employing distinct TMDs 
in the two bilayers, such as MoSe$_2$-WSe$_2$ for one and MoS$_2$-WS$_2$ for the other. 
We note that the symmetry breaking has the additional benefit of enabling differentiation between  
the light emission originating from the upper and lower excitons.



The excitonic energy shifts, associated with the chemical potentials of the two layers $\mu_{1,2}$, 
serve as an experimental indicator for a TSF phase. Specifically, in 
the zero-temperature limit of the TSF phase, $\mu_{1,2}$ should undergo a discontinuous jump when the 
density of either layer crosses the point $n_2=2n_1$. The magnitude of this jump, comparable to the 
energy difference between trimer and dimer bound states, is expected to vanish in the 2SF phase.
While we are unable to calculate $\mu_{1,2}$ at sufficiently low temperatures, 
we present their high-$T$ dependence on $N_2-2N_1$ in Appendix \ref{app:chempot}. In practice, 
however, the small energy scale of the jump makes its observation challenging, particularly in 
the presence of unavoidable disorder. Recently, more direct evidence for superfluidity of 
interlayer excitons was established by measuring the quasi-long-range spatial coherence of their 
photoluminescence \cite{exciton-superfluidity}, an approach that could similarly be applied in our 
context. Finally, it is possible to identify the presence of trimers by measuring interlayer drag. 
This involves driving the upper excitons to drift via a laterally modulated electric field, generated 
by patterned top gates \cite{Unuchek2018,Shanks2021}. A corresponding change in the cloud dynamics 
of the lower excitons is then sought by monitoring their light emission.


\section{Discussion and outlook }
We have established the formation of a trimer bound state in a bilayer of antiparallel 
dipolar excitons. For the case of a 1:2 density ratio between the layers, we have found 
that although the system is well described as a liquid of bound trimers across the entire 
studied density range, the resulting low-temperature superfluid still undergoes a quantum 
phase transition. In the low-density TSF phase, superflow is carried exclusively by trimers. 
However, across the QPT, new channels involving superflow of individual excitons in the two 
layers open up in parallel to that of the trimers. The thermal disordering of this 2SF phase 
is particularly rich, occurring via a cascade of BKT transitions that proceed through 
a reentrant TSF or a single-layer superfluid phase. Our predictions are directly 
relevant to experiments probing excitons in TMD heterostructures.

\begin{figure*}[t]
    \centering
    \includegraphics[width=1.0\linewidth]{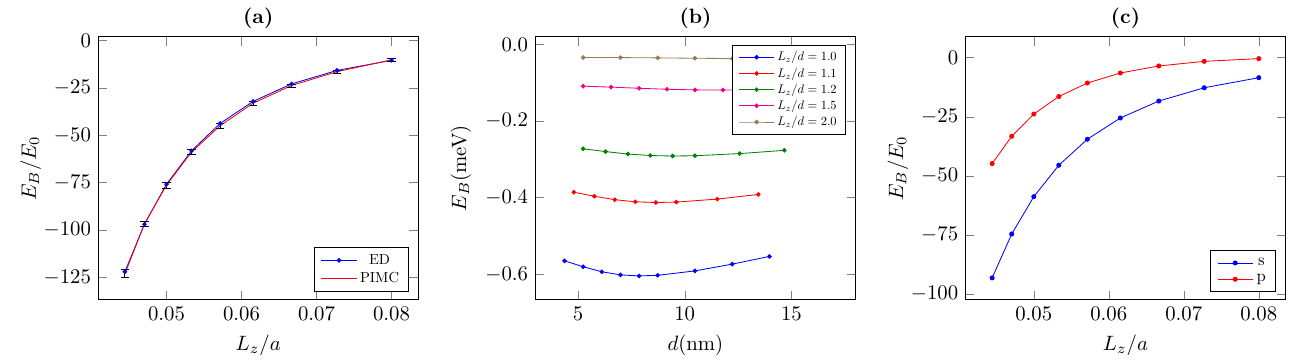}
    \captionsetup[subfigure]{labelformat=empty}
    \subfloat[\label{subfig:ED_QMC}]{}
    \subfloat[\label{subfig:BE_Lz_d}]{}
    \subfloat[\label{subfig:Dimer_s_p}]{}
    \vspace{-10pt}
    \caption{(a) Comparison between the trimer binding energy as calculated using exact diagonalization
    and PIMC. (b) The dependence of the trimer binding energy on $d$, the extent of the exciton along
    the $z$-axis, for various values of $L_z/d$, where $L_z$ is the interlayer separation between dipoles.
    The data are calculated for a TMD quadrilayer, similar in structure to the one shown in
    \cref{fig:Model}. (c) The binding energy of a dimer with angular momentum $l=0$ and $l=1$.}
    \label{fig:BindingEnergies}
\end{figure*}

While our primary focus was the 1:2 excitonic density ratio, we now briefly discuss the system's 
phase diagram as a function of the imbalance away from this point. We do so assuming 
that at zero temperature and zero imbalance, the system is in the TSF phase. When $N_1 < N_2 < 2N_1$, 
layer 2 lacks sufficient excitons to bind into trimers with all layer-1 excitons. Consequently, we 
expect the formation of $2N_1-N_2$ interlayer dimers coexisting alongside $N_2-N_1$ trimers, as other 
configurations, (e.g., involving "reversed" trimers with two layer-1 excitons) exhibit smaller binding 
energies. At zero temperature both trimers and dimers condense, breaking the full $U(1)\times  U(1)$ 
symmetry. However, for a small imbalance, the dimers are dilute, and thermal disordering of their 
phases leads to a BKT transition into a TSF phase at a critical temperature that scales linearly with 
the imbalance. For $N_2 > 2N_1$, the excess layer-2 excitons remain unbound and at low temperature 
condense in parallel to the trimers. The resulting 2SF phase also undergoes thermal BKT disordering 
into a TSF phase at a critical temperature that rises linearly with the imbalance.

Looking forward, our work opens several avenues for future research. By increasing the ratio 
of interaction to kinetic energy, it may be possible to stabilize a crystalline phase of trimers, 
analogous to phases seen in systems of simple dipoles \cite{Zoller_2007}. Furthermore, extending 
beyond the bilayer limit, engineering multilayer heterostructures with designed dipole orientations 
could pave the way for realizing novel composite quasi-particles and exotic correlated states. 
We leave these outstanding questions to future research.

\begin{acknowledgments}
We thank Ronen Rapaport and Hadar Steinberg for useful discussions. 
M.Z. acknowledges the support of the Council for Higher Education Scholarships Program 
for Outstanding Doctoral Students in Quantum Science and Technology. 
S.G. acknowledges support from the Israel Science Foundation (ISF) Grant no. 586/22 
and the US–Israel Binational Science Foundation (BSF) Grant no. 2020264. 
D.O. acknowledges support from the Israel Science Foundation (ISF) Grant no. 1975/24. 
\end{acknowledgments}

\section*{Data Availability}
The data that support the findings of this study are available at \cite{zenodo}.


\appendix

\section{Bound State Energies}
\label{app:boundstate}

We have used exact diagonalization, and PIMC simulations in the limit of zero temperature, to obtain the
ground state energy of the three-exciton Hamiltonian expressed in terms of
$\bm{\rho}_+=(\vb{r}_2+\vb{r}_3)/2L_z-\vb{r}_1/L_z$ and $\bm{\rho}_-=(\vb{r}_2-\vb{r}_3)/2L_z$
\begin{eqnarray}
\nonumber
 &&\hspace{-1.18cm}\mathcal{H}=-\frac{\hbar^2}{2m_X L_z^2} \left(\frac{3}{2}\nabla_+^2 + \frac{1}{2}\nabla_-^2\right)\\
 &&\hspace{-0.41cm}+\frac{D^2}{L_z^3}\left[\frac{1}{8\rho_-^3}-\sum_{\sigma=\pm}\frac{\abs{\bm{\rho}_++\sigma\bm{\rho}_-}^2 - 2}
  {\qty(\abs{\bm{\rho}_++\sigma\bm{\rho}_-}^2 + 1)^{5/2}}\right],
\label{eq:H_3body}
\end{eqnarray}
where we have omitted the trivial center of mass part. The agreement between the results, as shown in
Fig. \ref{subfig:ED_QMC}, attests to the validity of the PIMC calculation.

\begin{figure}[t]
    \centering
    \vspace{5pt}
    \includegraphics[width=0.93\linewidth]{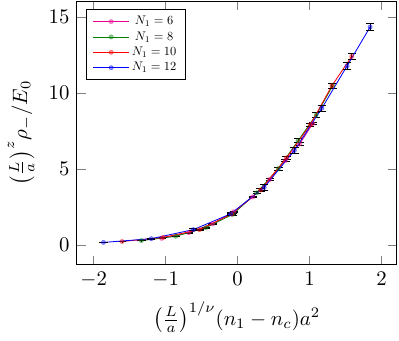}
    \vspace{15pt}
    \captionsetup[subfigure]{labelformat=empty}
    \caption{\centering Data collapse of the $\rho_-$ data using $z=2$, $\nu=0.61$ and $n_c=0.76$. }
    \label{fig:collapse}
\end{figure}
\begin{figure*}[t]
    \centering
    \includegraphics[width=1.0\linewidth]{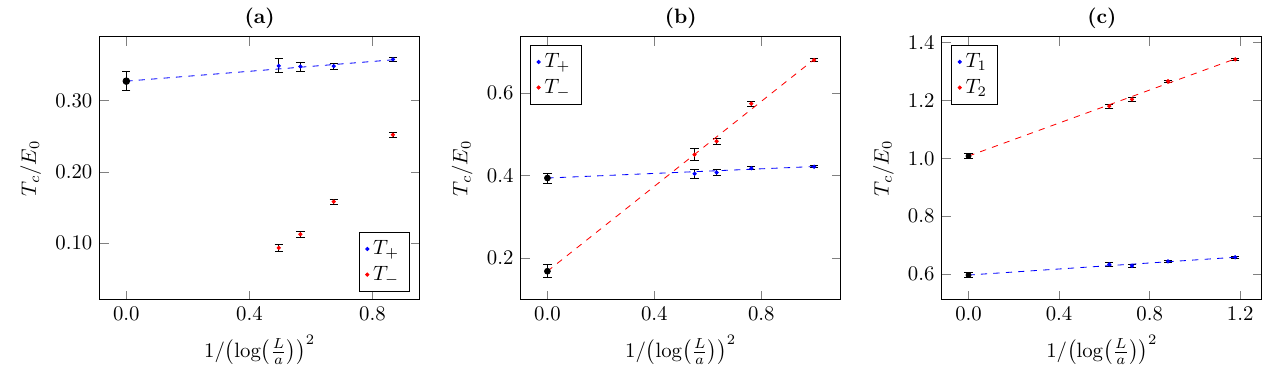}
    \captionsetup[subfigure]{labelformat=empty}
    \subfloat[\label{subfig:low}]{}
    \subfloat[\label{subfig:mid}]{}
    \subfloat[\label{subfig:high}]{}
    \caption{\centering Finite size scaling of the BKT critical temperatures for (a) $n_1=0.7/a^2$.
    (b) $n_1=0.81/a^2$. (c) $n_1=0.95/a^2$}
    \label{fig:FSS_Tc}
\end{figure*}

The binding energy of the trimer bound state depends on the distance $d$ between the layers
supporting a dipole, {\it i.e.} the extent of the exciton along the $z$-axis, and on the interlayer
separation $L_z$ between the dipoles. This is shown in Fig. \ref{subfig:BE_Lz_d} for a
MoSe$_2$-WSe$_2$ quadrilayer, similar in structure to the one shown in \cref{fig:Model}.
For fixed $L_z/d$, the kinetic term in Eq. (\ref{eq:H_3body}) scales as $1/d^2$, while the interaction
scales as $1/d$. Hence, for sufficiently small $d$ the kinetic energy dominates and leads, due to
delocalization effects, to a reduction in the binding energy with decreasing $d$. In the opposite
limit of large $d$, the interaction part dominates but the strength of the attractive
dipolar potential diminishes with $d$, and with it the binding energy.

We have also used exact diagonalization to calculate the binding energy of a two-exciton system,
described by the Hamiltonian
\begin{equation}
\begin{aligned}
  \mathcal{H}&=-\frac{\hbar^2}{m_X} \nabla^2 -{D^2}\frac{r^2 - 2L_z^2}{\qty(r^2 + L_z^2)^{5/2}}.
\end{aligned}
\label{eq:H_2body}
\end{equation}
The results for an s-wave and a p-wave dimer are depicted in Fig. \ref{subfig:Dimer_s_p}.
Evidently, the dimer ground state has zero angular momentum and its binding energy is smaller
than that of a trimer state by about 20\%, see Fig. \ref{subfig:ED_QMC}.

\section{$\rho_-$ Data Collapse}
\label{app:datacoll}

The zero temperature trimer superfluid phase
spontaneously breaks $\theta_+$. The gapless fluctuations in $\theta_+$ are equivalent to particle number fluctuations 
in the $N_+ = N_1 + 2N_2$ channel, corresponding to adding or removing trimers from the condensate. On the other hand, 
trimer-breaking excitations that change $N_- = 2N_1 - N_2$ and involve fluctuations in $\theta_-$ are gapped in the TSF.

As the density is increased, the gap in the $\theta_-$ mode decreases and eventually vanishes at $n_c$. The nature 
of the transition depends on whether there is a particle-hole symmetry that makes the gaps involved in increasing
and decreasing $N_-$ vanish symmetrically. In such a case the transition is relativistic with a dynamical exponent $z=1$, 
and belongs to the 3D XY universality class. In the absence of such a symmetry, as is relevant to our case, the transition 
is expected to be a non-relativistic, $z=2$, superfluid transition.

With the above discussion in mind, we can write a scaling form for the stiffness
\begin{equation}
\frac{\rho_-(n_1,L)}{E_0} = \left(\frac{L}{a}\right)^{-z} f\left[\left(\frac{L}{a}\right)^{1/\nu} (n_1-n_c) a^2\right] ,
\end{equation}
as a function of the rescaled parameter $L^{1/\nu} (n_1-n_c)$. A curve fit gives our numerical estimates, $\nu=0.61 \pm 0.04$ 
and $n_c = 0.758 \pm 0.006$, with $z=2$, as shown in the curve collapse plot \cref{fig:collapse}. We note that our estimate 
of $\nu$ is smaller than the 3D XY result $\nu_{\text{3DXY}}=0.67$, yet larger than the mean field result of $\nu_{\text{MF}}=0.5$. 
We conjecture that the discrepancy stems from the larger finite-size effects observed in $\rho_-$, as discussed below.

\section{Finite Size Scaling of Critical Temperatures}
\label{app:Tcscale}

The finite size scaling of the critical temperatures of the BKT transitions, deduced from the
corresponding stiffnesses, is shown in Fig \ref{fig:FSS_Tc} for three densities across the
phase diagram. The data is consistent with $T_c \propto 1/(\log L)^2$ scaling, where $L$ is the system size, as expected from BKT theory.

We find that the critical temperature $T_+$, calculated from $\rho_+$ and associated with the TSF-Normal transition,
exhibits only weak dependence on $L$. The same is true for the critical temperature $T_1$ of the transition between
the 2SF and SF2 phases, as calculated from $\rho_1$. In contrast, the data for $\rho_-$ and $\rho_2$ show considerably
larger finite size effects. For $n_1=0.7/a^2$ our finite-size results for $T_-$, shown in Fig. \ref{subfig:low},
are consistent with the vanishing of the low temperature $\rho_-$, see Fig \ref{subfig:FSS},
and indicate the presence of a zero-temperature TSF.
Finite size effects in $\rho_-$ are large at $n_1=0.81/a^2$,
where the scaling shown in Fig. \ref{subfig:mid} clearly indicates that $T_+ > T_-$ in the thermodynamic
limit, although the reverse is true for all the finite systems that we have simulated. This
demonstrates a thermal reentrance of the TSF phase close to the critical density. At higher densities,
as depicted in Fig. \ref{subfig:high}, $T_2$ shows more significant changes with $L$ in comparison to $T_1$, but
stays consistently above it thereby establishing the SF2 phase.
\begin{figure}[t!!!]
    \centering
    \includegraphics[width=0.98\linewidth]{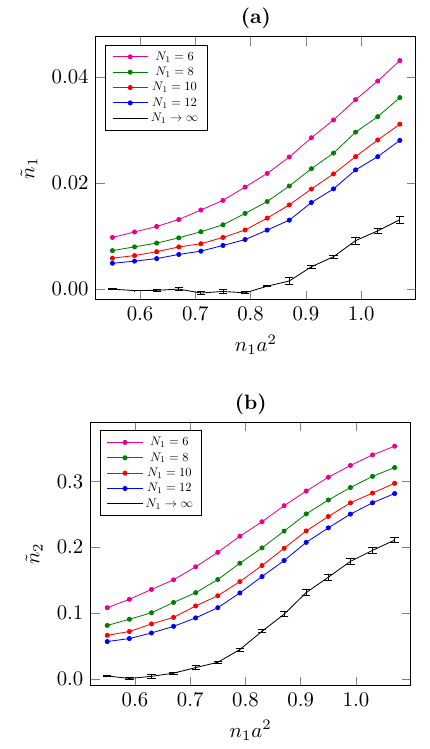}
    \captionsetup[subfigure]{labelformat=empty}
    \subfloat[\label{subfig:sf1}]{}
    \subfloat[\label{subfig:sf2}]{}
    \caption{(a) Finite size scaling of the superfluid fraction in layer 1 as a
    function density. (b) Finite size scaling of the superfluid fraction in layer 2. The black
    lines depict the extrapolation to the thermodynamic limit.}
    \label{fig:superfluid_fraction}
\end{figure}

\begin{figure*}[t]
    \centering
    \includegraphics[width=1.0\linewidth]{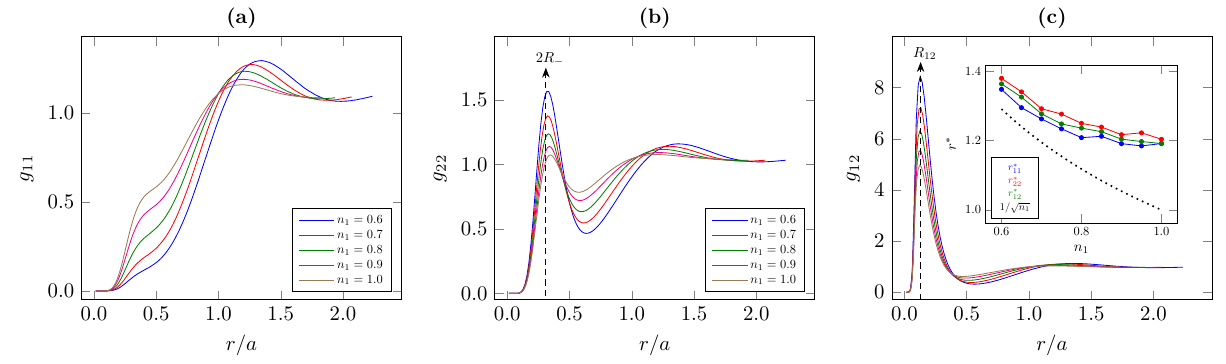}
    \captionsetup[subfigure]{labelformat=empty}
    \subfloat[\label{subfig:g11}]{}
    \subfloat[\label{subfig:g22}]{}
    \subfloat[\label{subfig:g12}]{}
    \caption{The density-density correlation functions in the zero-temperature limit for various
    densities $n_1$. (a) The peak in $g_{11}$ resides at $1/\sqrt{n_1}$, while the shoulder appears
    at a separation slightly larger than the trimer diameter $2R_-$, which is also where $g_{22}$
    exhibits a peak, as shown in (b). (c) The interlayer correlations peaks at $R_{12}$, where
    the probability distribution $|\psi(\bf{r}_1-\bf{r}_2)|^2$ attains its maximum. The inset depicts
    the positions $r_{\alpha\beta}^{*}$ of the first maximum in $g_{11}$ and the second maximum
    of $g_{12}$ and $g_{22}$ as a function of $n_1$.}
    \label{fig:correlations}
\end{figure*}

\section{Superfluid Fractions}
\label{app:sffrac}

The superfluid fraction $\tilde{n}_\alpha$ in layer $\alpha$ is defined in terms of the one-particle
equal-time Green's functions as
\begin{equation}
\begin{aligned}
    \tilde{n}_{\alpha} = \frac{1}{A^2}\int d^2 \vb{r} \int d^2 \vb{r}' \expval{\hat{\psi}^{\dagger}_{\alpha}(\vb{r}) \hat{\psi}_{\alpha}(\vb{r}')},
\end{aligned}
\label{eq:superfluid_fraction}
\end{equation}
where $A$ is the system area. In the TSF phase, both $\theta_{1}$ and $\theta_2$ are disordered
and consequently $\tilde{n}_{1,2}=0$. Conversely, in the ordered 2SF phase both $\tilde{n}_{1, 2} \neq 0$.
In Fig. \ref{fig:superfluid_fraction} we use the expected $1/N_1$ scaling of the superfluid fractions to
extrapolate the finite size data to the thermodynamic limit $N_1\rightarrow\infty$. We find that
the extrapolated $\tilde{n}_1$ clearly vanishes for $n_1<0.79/a^2$ and turns positive at higher
densities, consistent with the position of the QPT as deduced from the behavior of $\rho_-$.
The extrapolated $\tilde{n}_2$ follows a similar trend but does not fully vanish in the range
$0.65<n_1 a^2<0.79$. It appears that the layer-2 data is more susceptible to finite size effects,
as was already noted in the context of the scaling of the critical temperatures in the previous section.

\section{Density-Density Correlation functions}
\label{app:dencorr}

Fig. \ref{fig:correlations} depicts the density-density correlation functions $g_{\alpha\beta}$
in the zero-temperature limit, as $n_1$ is increased across the QPT from the TSF to the 2SF phase.
The positions of the peaks in $g_{22}$ and $g_{12}$ are close to $2R_-$ and $R_{12}$ (the
positions of the maxima in the single trimer probability distributions $|\psi_-|^2$ and $|\psi_{12}|^2$, respectively),
and increase marginally with density. Concomitantly, their height decrease.
This shows that the trimers largely maintain their spatial structure across
the QPT but that independent motion of excitons is gradually established. The fact that
$R_{12} < R_-$, reflects our observation that the top-layer exciton is typically found closer
to one of the two bottom-layer excitons and switches between them as the PIMC configuration
evolves in imaginary time.

Further insights can be gleaned from the behavior of the first maximum of $g_{11}$ and the second
maximum of $g_{12}$ and $g_{22}$. Their positions $r_{\alpha\beta}^{*}$, shown in the inset of
Fig. \ref{subfig:g12}, provide a measure for the most probable distance between nearest neighboring
trimers. The fact that these peaks continue to exist in the 2SF phase and to follow closely each other
is an additional evidence that the trimers do not fall apart when the system exits the TSF phase.
The data show that the critical density marks a transition from a $1/\sqrt{n_1}$ scaling of
$r_{\alpha\beta}^{*}$ in the dilute limit where the trimers are largely independent, to
$r_{\alpha\beta}^{*}$ which change little with density in the 2SF phase where trimers can
not be pushed closer to each other due to their mutual repulsion.

Finally, as density increases, a bump develops in $g_{11}$ at slightly larger separation than the trimer
diameter $2R_-$. This indicates that trimers spend increasingly more time in the vicinity
of each other. Such events allow for more effective exchange of excitons between trimers and
establish windings by independent excitons around the sample that lead to the 2SF phase.

\section{Exciton Winding}
\label{app:windings}

To gain insights into the nature of exciton winding
we include two movies in \cite{suppmat} showing the imaginary time evolution of the exciton configuration over $\tau\in[0,\beta]$ at a temperature $T=0.1E_0$.
Red and blue colors denote layer-1 and layer-2 particles, respectively. At each time slice we group the particles into trios by
minimizing the sum over all trios of the distance between the layer-1 particle and its layer-2 partners within the trio.
To help follow the evolution of the windings we display
next to each particle its instantaneous winding number, which increases or decreases by one when it crosses the dashed boundary from
left to right or from  right to left, respectively. The title depicts the imaginary time,
the sum total of the windings $W_{1,2}$
of particles in each layer, and the relative winding $W_- = 2W_1-W_2$. The variance of $W_-(\tau=\beta)$ determines $\rho_-$.

The movie TSF.mp4 \cite{suppmat} follows an $n_1 a^2=0.7$
system that is in a TSF phase. Clearly, the particles are bound into trimers, and crossings of the
boundary typically involve an entire trimer. During such events $W_-$ becomes momentarily nonzero, but then vanishes again
when the trimer completes its crossing. Alternatively, two trimers may come together from both sides of the boundary and
exchange a pair of particles between them. These events also maintain a zero $W_-$, which is the hallmark of a TSF.

The movie 2SF.mp4 \cite{suppmat}
follows an $n_1 a^2=0.95$ system that is in a 2SF phase. One observes that the particles still spend most of the time within
well-defined trimers that follow the dynamics described above. However, there is an increased number of cases where a layer-2
particle lingers away from its partners. Instances of this type allow for a new type of winding events where such a particle
crosses the boundary and joins a new trimer, without its old partners crossing the boundary as well. This is followed by a  global
rearrangement of the particles into a new configuration of trimers. The net result is a change in $W_-$ that can establish
non-vanishing $\rho_-$ and hence a 2SF phase.

\begin{figure}[t!]
    \centering
    \includegraphics[width=0.91\linewidth]{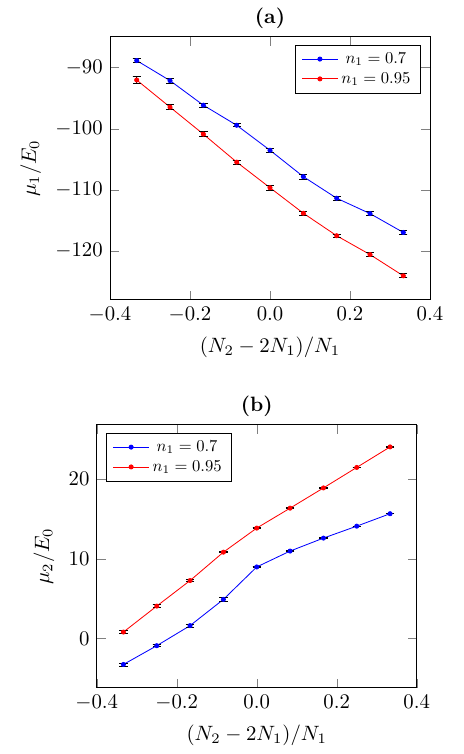}
    \captionsetup[subfigure]{labelformat=empty}
    \subfloat[\label{subfig:mu1}]{}
    \subfloat[\label{subfig:mu2}]{}
    \caption{The chemical potentials of the two layers as a function of the relative deviation from
    the point $N_2=2N_1$. The data are shown for a system with $N_1=12$ at the two indicated densities.
    (a) $\mu_1$ at $T=2.5E_0$, (b) $\mu_2$ at $T=1.5E_0$.}
    \label{fig:chemical-potentials}
\end{figure}

\section{Chemical Potentials}
\label{app:chempot}

We calculate the finite temperature chemical potentials of both layers
$\mu_{\alpha}=F(N_{\alpha}+1, T)-F(N_{\alpha}, T)$ as a function of the relative deviation $(N_2-2N_1)/N_1$
from the $1:2$ density ratio. We control the deviation by fixing $N_1$ and
varying $N_2$. We consider two densities $n_1=0.7$ and $n_1=0.95$, where the low-temperature phase of
the system is a TSF and 2SF, respectively. However, we were unable to obtain reliable $\mu_{1,2}$ data
within the superfluid phases. Therefore, Fig. \ref{fig:chemical-potentials} depicts results for $\mu_1$ 
at $T=2.5E_0$ and for $\mu_2$ at $T=1.5E_0$, where the system is in its normal state.

As noted in the Discussion, when $N_1 < N_2 < 2N_1$, $2N_1-N_2$ interlayer dimers, each comprising 
a pair of excitons from the two layers, are expected to form alongside the trimers. Under such 
conditions, the addition of an exciton to layer 1 converts a trimer into a pair of dimers. In the 
dilute limit this leads to a reduction of the energy by $2E_D - E_T \simeq 60E_0$, where $E_D$ and 
$E_T$ are the dimer and trimer binding energies, respectively. The gain in free energy is further 
enhanced by the attractive interaction between the added exciton and the dipoles in layer 2. Its 
effect outweighs that of the repulsion within layer 1 due to the higher density in layer 2. 
Consequently, $\mu_1$ is expected to be large and negative, increasing in magnitude as $N_2$ 
increases. The data presented in Fig. \ref{subfig:mu1} align with this expectation.

Conversely, the addition of an exciton to layer 2 of a system with $N_1 < N_2 < 2N_1$ converts a 
dimer into a trimer and reduces the energy by $E_T - E_D \simeq 20E_0$. However, the interaction 
of the added exciton with the other dipoles modifies this result. The dominant role is played by 
the repulsion from the other layer-2 excitons, which diminishes the energetic gain and leads to a 
negative $\mu_2$ whose magnitude is smaller than $E_T - E_D$ and decreases with $N_2$, 
as shown in Fig. \ref{subfig:mu2}.

Excess excitons in layer $2$ remain unbound for $N_2 > 2N_1$. Adding a layer-1 exciton binds two 
layer-2 excitons into a trimer, which in the dilute limit reduces the energy by $E_T \simeq 100E_0$. 
The attraction from layer-2 excitons further reduces the energy, thus making $\mu_1$ more negative 
with increasing $N_2$, see Fig. \ref{subfig:mu1}. Since additional layer-2 dipoles do not bind when 
$N_2>N_1$ and are primarily affected by intralayer repulsion, $\mu_2$ is positive and increases with 
$N_2$ in this parameter range, as demonstrated by Fig. \ref{subfig:mu2}.

We expect that a TSF leads to a discontinuous jump in $\mu_1$ and $\mu_2$ at $T=0$ and $N_2 = 2N_1$, 
which in the dilute limit approach $-2(E_T-E_D)$ and $E_T-E_D$, respectively. The jump gets smoother 
as temperature is increased. Although we are unable to reach the required low temperatures, we note 
that $\mu_2$ exhibits a kink at $N_2=2N_1$  for $n_1=0.7$.

\begin{figure*}[t]
    \centering
    \includegraphics[width=\linewidth]{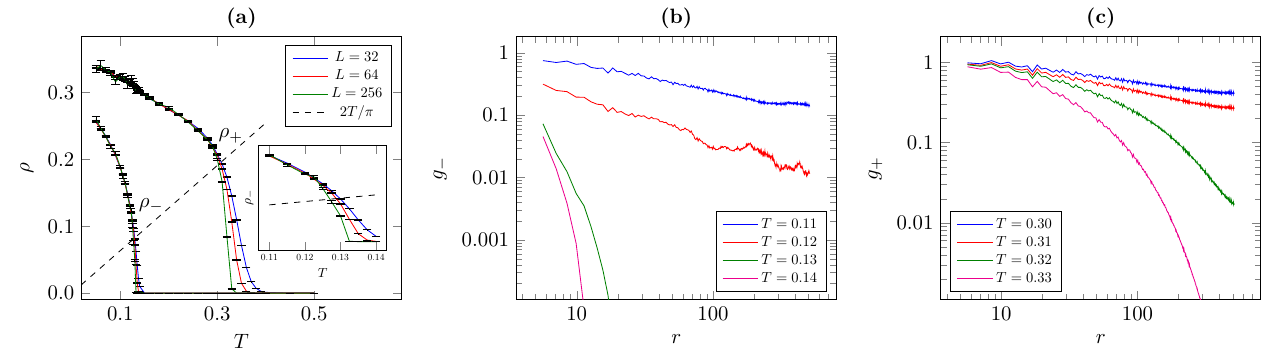}
    \captionsetup[subfigure]{labelformat=empty}
    \subfloat[\label{subfig:XY_mid_stiff}]{}
    \subfloat[\label{subfig:XY_mid_g-}]{}
    \subfloat[\label{subfig:XY_mid_g+}]{}
    \caption{(a) The temperature dependence of the stiffnesses, $\rho_-$ and $\rho_+$, associated with
    the phases $\theta_-/5$ and $\theta_+$, for $n_1=1.05$. Both develop a jump when crossing the $2T/\pi$
    line, as the system size increases. The inset zooms on the vicinity of the jump in $\rho_-$.
    (b) and (c) The $\theta_-$ and $\theta_+$ two-point correlations as a function of separation for a system with $L=1024$ at
    various temperatures. }
    \label{fig:XY_mid}
\end{figure*}

\vspace{1cm}
\section{Effective XY Model}
\label{app:xy}

We employ classical Monte Carlo worm-algorithm simulations to study the effective XY model
\begin{widetext}
\begin{eqnarray}
    \label{eq:effXYsup}
    \nonumber
    {\cal H}&=&-\sum_\mu \left\{(n_1-n_c)\left[\cos(\nabla_\mu\theta_1)
    + 2\cos(\nabla_\mu\theta_2)\right]
    +\frac{n_c} {3}\cos\left[\nabla_\mu(\theta_1+2\theta_2)\right]\right\} \\
    &=&-\sum_\mu \left\{(n_1-n_c)\left[\cos\left(\nabla_\mu\frac{\theta_+ + 2\theta_-}{5}\right)
    + 2\cos\left(\nabla_\mu\frac{2\theta_+ - \theta_-}{5}\right) \right]
    +\frac{n_c} {3}\cos\left(\nabla_\mu \theta_+\right)\right\}
\end{eqnarray}
on an $L\times L$ lattice. To this end, we use the loop representation
of the partition function as a sum over configurations of closed current loops in both layers
\begin{equation}
\begin{aligned}
      \mathcal{Z} = \sum\limits_{\qty{\sum\limits_{\mu}\nabla_{\mu}j^{\alpha}_{\mu} = 0}}\prod\limits_{\mu, \vb{r}}\qty[\sum\limits_{m=-\infty}^{\infty}I_m\qty(\frac{n_c}{3T})I_{(j^1_{\mu} - m)}\qty(\frac{n_1-n_c}{T})I_{(j^2_{\mu} - 2m)}\qty(2\frac{n_1-n_c}{T})].
\end{aligned}
\label{eq:XY_loop}
\end{equation}
\end{widetext}
Here, $T$ is the temperature, $I_m$ are the modified Bessel functions of the first kind, and
$j^{\alpha}_{\mu}({\bm r})$ is the integer current variable associated with the bond
emanating from site ${\bm r}$ on layer $\alpha$ along the $\mu$ direction.
Within this representation the stiffnesses are calculated as
\begin{equation}
\begin{aligned}
  \rho_{\alpha\beta} = T\expval{\qty(\sum\limits_{\mu,\vb{r}}j^{\alpha}_{\mu})
  \qty(\sum\limits_{\mu,\vb{r}}j^{\beta}_{\mu})}.
\end{aligned}
\end{equation}
We set $n_c = 1$ and consider two regimes, one near the critical density at $n_1 = 1.05$,
and the other further away from the critical density at $n_1 = 2$.

\begin{figure*}[t]
    \centering
    \includegraphics[width=\linewidth]{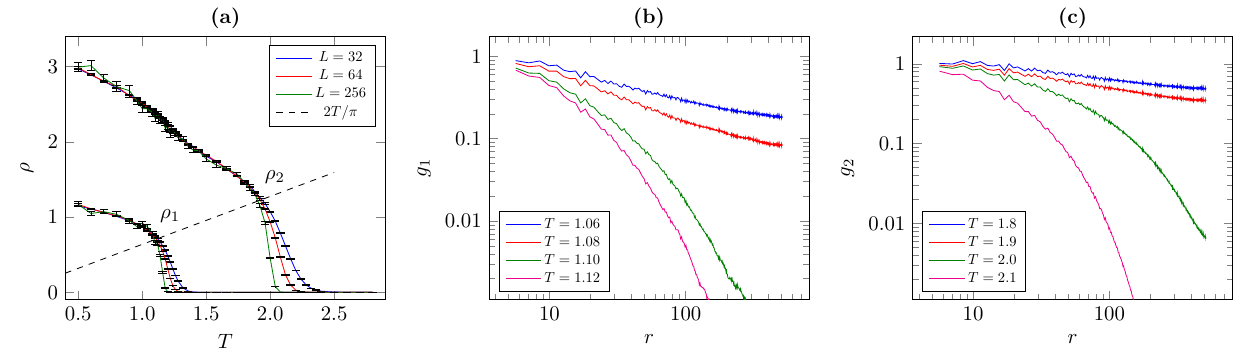}
    \captionsetup[subfigure]{labelformat=empty}
    \subfloat[\label{subfig:XY_high_stiff}]{}
    \subfloat[\label{subfig:XY_high_g1}]{}
    \subfloat[\label{subfig:XY_high_g2}]{}
    \vspace{-3pt}
    \caption{(a) The temperature dependence of the stiffnesses, $\rho_1$ and $\rho_2$, associated with
    the phases $\theta_1$ and $\theta_2$, for $n_1=2$. Both develop a jump when crossing the $2T/\pi$
    line, as the system size increases. (b) and (c) The $\theta_1$ and $\theta_2$ two-point correlations
    as a function of separation for a system with $L=1024$ at various temperatures.}
    \label{fig:XY_high}
\end{figure*}

\subsection{The 2SF-TSF-Normal Transitions}
At $n_1 = 1.05$, the bare stiffness associated with the first two terms in Eq. (\ref{eq:effXYsup})
is small and the trimer term $\sum_\mu \cos(\nabla_{\mu}\theta_+)$ dominates. Consequently,
the transition out of the low temperature 2SF phase is driven by the proliferation of $\theta_-$
vortices. These are constrained to obey $m_1+2m_2=0$, where $m_\alpha$ is the vorticity in layer $\alpha$,
since $\theta_+$ retains its QLRO through the transition to the TSF phase.

There are two possible scenarios for the disordering of $\theta_-$ \cite{Lee_1985,Carpenter_1989}.
The first involves $\theta_-$ vortices with a $10\pi$ phase twist ($m_1=2$, $m_2=-1$) caused by
the $\theta_-/5$ term in Eq. (\ref{eq:effXYsup}). Their proliferation induces a BKT transition
with a jump of $2T_c/\pi$ in the stiffness $\rho_-=4\rho_1+\rho_2-4\rho_{12}$ associated with the
phase $\theta_-/5$. Alternatively, the model may undergo an Ising-like transition into a nematic
phase, followed by a BKT transition due to proliferation of $\theta_-$ vortices with a $5\pi$
phase twist ($m_1=1$, $m_2=-1/2$). Such a transition would be accompanied by an $8T_c/\pi$ jump
in $\rho_-$. Fig. \ref{subfig:XY_mid_stiff} gives evidence in favor of the first scenario by
showing that as $L$ increases, $\rho_-$ develops a jump when crossing the $2T/\pi$ line.
Concomitantly, as shown by Fig. \ref{subfig:XY_mid_g-}, the decay of the $\theta_-$ correlation
function $g_-(r) = \expval{e^{i[\theta_-(r) - \theta_-(0)]}}$ changes from power-law to exponential
when the temperature is raised through the location of the $\rho_-$ jump in Fig. \ref{subfig:XY_mid_stiff}.
As the temperature is further increased, QLRO in $\theta_+$ is lost due to proliferation of $2\pi$
vortices in $\theta_+$, see Fig. \ref{subfig:XY_mid_stiff} and Fig. \ref{subfig:XY_mid_g+}.

\vspace{10pt}
\subsection{The 2SF-SF2-Normal Transitions}

At $n_1 = 2$, the dominant term in Eq. (\ref{eq:effXYsup}) is $\sum_\mu\cos(\nabla_{\mu}\theta_2)$. 
Thus, $\theta_2$ retains QLRO across the low-temperature 2SF-SF2 BKT transition, while $\theta_1$ 
becomes disordered by proliferation of $2\pi$ vortices. The transition is identified by the onset of 
a jump in $\rho_1$ as it crosses the $2T/\pi$ line and by the accompanying exponential decay of the 
two-point $\theta_1$ correlations $g_1(r) = \langle e^{i[\theta_1(r) - \theta_1(0)]}\rangle$, 
see Fig. \ref{subfig:XY_high_stiff} and Fig. \ref{subfig:XY_high_g1}. The phase $\theta_2$ undergoes 
a similar disordering transition at a higher temperature due to proliferation of $2\pi$ vortices, 
as shown by Fig. \ref{subfig:XY_high_stiff} and Fig. \ref{subfig:XY_high_g2}. The system then exits 
the SF2 phase into the normal state.

\bibliography{TSF_PRB.bib}

@ARTICLE{Zimmerman_2022,
       author = {{Zimmerman}, Michal and {Rapaport}, Ronen and {Gazit}, Snir},
        title = "{Collective interlayer pairing and pair superfluidity in vertically stacked layers of dipolar excitons}",
      journal = {Proc. Natl. Acad. Sci. U.S.A.},
         year = 2022,
        month = jul,
       volume = {119},
       number = {30},
          eid = {e2205845119},
        pages = {e2205845119},
          doi = {10.1073/pnas.2205845119}
}

@article{Paik_24,
author = {Eunice Paik and Long Zhang and Kin Fai Mak and Jie Shan and Hui Deng},
journal = {Adv. Opt. Photon.},
number = {4},
pages = {1064--1132},
publisher = {Optica Publishing Group},
title = {Excitons and polaritons  in two-dimensional transition metal dichalcogenides: a tutorial},
volume = {16},
month = {Dec},
year = {2024},
url = {https://opg.optica.org/aop/abstract.cfm?URI=aop-16-4-1064},
doi = {10.1364/AOP.504035},
}

@article{Zoller_2007,
  title = {Strongly Correlated {2D} Quantum Phases with Cold Polar Molecules: Controlling the Shape of the Interaction Potential},
  author = {B\"uchler, H. P. and Demler, E. and Lukin, M. and Micheli, A. and Prokof'ev, N. and Pupillo, G. and Zoller, P.},
  journal = {Phys. Rev. Lett.},
  volume = {98},
  issue = {6},
  pages = {060404},
  numpages = {4},
  year = {2007},
  month = {Feb},
  publisher = {American Physical Society},
  doi = {10.1103/PhysRevLett.98.060404},
  url = {https://link.aps.org/doi/10.1103/PhysRevLett.98.060404}
}

@article{Ferlaino_2012,
  title = {Bose-{E}instein Condensation of Erbium},
  author = {Aikawa, K. and Frisch, A. and Mark, M. and Baier, S. and Rietzler, A. and Grimm, R. and Ferlaino, F.},
  journal = {Phys. Rev. Lett.},
  volume = {108},
  issue = {21},
  pages = {210401},
  numpages = {5},
  year = {2012},
  month = {May},
  publisher = {American Physical Society},
  doi = {10.1103/PhysRevLett.108.210401},
  url = {https://link.aps.org/doi/10.1103/PhysRevLett.108.210401}
}

@article{Lev_2011,
  title = {Strongly Dipolar {B}ose-{E}instein Condensate of Dysprosium},
  author = {Lu, Mingwu and Burdick, Nathaniel Q. and Youn, Seo Ho and Lev, Benjamin L.},
  journal = {Phys. Rev. Lett.},
  volume = {107},
  issue = {19},
  pages = {190401},
  numpages = {5},
  year = {2011},
  month = {Oct},
  publisher = {American Physical Society},
  doi = {10.1103/PhysRevLett.107.190401},
  url = {https://link.aps.org/doi/10.1103/PhysRevLett.107.190401}
}

@article{Tilman_2005,
  title = {Bose-{E}instein Condensation of Chromium},
  author = {Griesmaier, Axel and Werner, J\"org and Hensler, Sven and Stuhler, J\"urgen and Pfau, Tilman},
  journal = {Phys. Rev. Lett.},
  volume = {94},
  issue = {16},
  pages = {160401},
  numpages = {4},
  year = {2005},
  month = {Apr},
  publisher = {American Physical Society},
  doi = {10.1103/PhysRevLett.94.160401},
  url = {https://link.aps.org/doi/10.1103/PhysRevLett.94.160401}
}

@article{Butov_2004,
doi = {10.1088/0953-8984/16/50/R02},
url = {https://dx.doi.org/10.1088/0953-8984/16/50/R02},
year = {2004},
month = {dec},
publisher = {},
volume = {16},
number = {50},
pages = {R1577},
author = {Butov, L V},
title = {Condensation and pattern formation in cold exciton gases in coupled quantum wells},
journal = {J. Phys. Condens. Matter}
}

@article{Rapaport_2007,
doi = {10.1088/0953-8984/19/29/295207},
url = {https://dx.doi.org/10.1088/0953-8984/19/29/295207},
year = {2007},
month = {jun},
publisher = {},
volume = {19},
number = {29},
pages = {295207},
author = {Rapaport, Ronen and Chen, Gang},
title = {Experimental methods and analysis of cold and dense dipolar exciton fluids},
journal = {J. Phys. Condens. Matter}
}

@article{Gossard_2021,
  title = {Attractive and repulsive dipolar interaction in bilayers of indirect excitons},
  author = {Choksy, D. J. and Xu, Chao and Fogler, M. M. and Butov, L. V. and Norman, J. and Gossard, A. C.},
  journal = {Phys. Rev. B},
  volume = {103},
  issue = {4},
  pages = {045126},
  numpages = {10},
  year = {2021},
  month = {Jan},
  publisher = {American Physical Society},
  doi = {10.1103/PhysRevB.103.045126},
  url = {https://link.aps.org/doi/10.1103/PhysRevB.103.045126}
}

@article{Rapaport_2019,
  title = {Attractive Dipolar Coupling between Stacked Exciton Fluids},
  author = {Hubert, Colin and Baruchi, Yifat and Mazuz-Harpaz, Yotam and Cohen, Kobi and Biermann, Klaus and Lemeshko, Mikhail and West, Ken and Pfeiffer, Loren and Rapaport, Ronen and Santos, Paulo},
  journal = {Phys. Rev. X},
  volume = {9},
  issue = {2},
  pages = {021026},
  numpages = {12},
  year = {2019},
  month = {May},
  publisher = {American Physical Society},
  doi = {10.1103/PhysRevX.9.021026},
  url = {https://link.aps.org/doi/10.1103/PhysRevX.9.021026}
}

@article{Lahaye_2009,
doi = {10.1088/0034-4885/72/12/126401},
url = {https://dx.doi.org/10.1088/0034-4885/72/12/126401},
year = {2009},
month = {nov},
publisher = {},
volume = {72},
number = {12},
pages = {126401},
author = {Lahaye, T and Menotti, C and Santos, L and Lewenstein, M and Pfau, T},
title = {The physics of dipolar bosonic quantum gases},
journal = {Rep. Prog. Phys.}
}

@article{Chomaz_2022,
   title={Dipolar physics: a review of experiments with magnetic quantum gases},
   volume={86},
   ISSN={1361-6633},
   url={http://dx.doi.org/10.1088/1361-6633/aca814},
   DOI={10.1088/1361-6633/aca814},
   number={2},
   journal={Rep. Prog. Phys.},
   publisher={IOP Publishing},
   author={Chomaz, Lauriane and Ferrier-Barbut, Igor and Ferlaino, Francesca and Laburthe-Tolra, Bruno and Lev, Benjamin L and Pfau, Tilman},
   year={2022},
   month=dec, pages={026401} }

@article{Ferlaino_2019,
  title = {Long-Lived and Transient Supersolid Behaviors in Dipolar Quantum Gases},
  author = {Chomaz, L. and Petter, D. and Ilzh\"ofer, P. and Natale, G. and Trautmann, A. and Politi, C. and Durastante, G. and van Bijnen, R. M. W. and Patscheider, A. and Sohmen, M. and Mark, M. J. and Ferlaino, F.},
  journal = {Phys. Rev. X},
  volume = {9},
  issue = {2},
  pages = {021012},
  numpages = {12},
  year = {2019},
  month = {Apr},
  publisher = {American Physical Society},
  doi = {10.1103/PhysRevX.9.021012},
  url = {https://link.aps.org/doi/10.1103/PhysRevX.9.021012}
}

@article{Modugno_2019,
  title = {Observation of a Dipolar Quantum Gas with Metastable Supersolid Properties},
  author = {Tanzi, L. and Lucioni, E. and Fam\`a, F. and Catani, J. and Fioretti, A. and Gabbanini, C. and Bisset, R. N. and Santos, L. and Modugno, G.},
  journal = {Phys. Rev. Lett.},
  volume = {122},
  issue = {13},
  pages = {130405},
  numpages = {6},
  year = {2019},
  month = {Apr},
  publisher = {American Physical Society},
  doi = {10.1103/PhysRevLett.122.130405},
  url = {https://link.aps.org/doi/10.1103/PhysRevLett.122.130405}
}

@article{Tilman_2019,
  title = {Transient Supersolid Properties in an Array of Dipolar Quantum Droplets},
  author = {B\"ottcher, Fabian and Schmidt, Jan-Niklas and Wenzel, Matthias and Hertkorn, Jens and Guo, Mingyang and Langen, Tim and Pfau, Tilman},
  journal = {Phys. Rev. X},
  volume = {9},
  issue = {1},
  pages = {011051},
  numpages = {7},
  year = {2019},
  month = {Mar},
  publisher = {American Physical Society},
  doi = {10.1103/PhysRevX.9.011051},
  url = {https://link.aps.org/doi/10.1103/PhysRevX.9.011051}
}

@article{Ceperley_1995,
  title = {Path integrals in the theory of condensed helium},
  author = {Ceperley, D. M.},
  journal = {Rev. Mod. Phys.},
  volume = {67},
  issue = {2},
  pages = {279--355},
  numpages = {0},
  year = {1995},
  month = {Apr},
  publisher = {American Physical Society},
  doi = {10.1103/RevModPhys.67.279},
  url = {https://link.aps.org/doi/10.1103/RevModPhys.67.279}
}

@article{Boninsegni_2006,
  title = {Worm Algorithm for Continuous-Space Path Integral {M}onte {C}arlo Simulations},
  author = {Boninsegni, Massimo and Prokof'ev, Nikolay and Svistunov, Boris},
  journal = {Phys. Rev. Lett.},
  volume = {96},
  issue = {7},
  pages = {070601},
  numpages = {4},
  year = {2006},
  month = {Feb},
  publisher = {American Physical Society},
  doi = {10.1103/PhysRevLett.96.070601},
  url = {https://link.aps.org/doi/10.1103/PhysRevLett.96.070601}
}

@unpublished{suppmat,
note = {The Supplemental Material at \url{https://doi.org/10.1103/zn5r-2vrb} includes two movies demonstrating particle windings in the TSF and 2SF phases.}
}

@article{Kuklov_2004,
	author = {Kuklov, Anatoly and Prokof'ev, Nikolay and Svistunov, Boris},
	date = {2004/01/22/},
	day = {22},
	doi = {10.1103/PhysRevLett.92.030403},
	id = {10.1103/PhysRevLett.92.030403},
	j1 = {PRL},
	journal = {Phys. Rev. Lett.},
	month = {01},
	number = {3},
	pages = {030403--},
	publisher = {American Physical Society},
	title = {Superfluid-Superfluid Phase Transitions in a Two-Component {B}ose-{E}instein Condensate},
	url = {https://link.aps.org/doi/10.1103/PhysRevLett.92.030403},
	volume = {92},
	year = {2004}
}

@article{Pollock_1987,
  title = {Path-integral computation of superfluid densities},
  author = {Pollock, E. L. and Ceperley, D. M.},
  journal = {Phys. Rev. B},
  volume = {36},
  issue = {16},
  pages = {8343},
  numpages = {10},
  year = {1987},
  month = {December},
  publisher = {American Physical Society},
  doi = {10.1103/PhysRevB.36.8343},
  url = { https://doi.org/10.1103/PhysRevB.36.8343}
}

@article{Lee_1985,
	author = {Lee, D. H. and Grinstein, G.},
	date = {1985/07/29/},
	day = {29},
	doi = {10.1103/PhysRevLett.55.541},
	id = {10.1103/PhysRevLett.55.541},
	j1 = {PRL},
	journal = {Phys. Rev. Lett.},
	month = {07},
	number = {5},
	pages = {541--544},
	publisher = {American Physical Society},
	title = {Strings in two-dimensional classical {XY} models},
	url = {https://link.aps.org/doi/10.1103/PhysRevLett.55.541},
	volume = {55},
	year = {1985}
}

@article{Carpenter_1989,
	author = {D B Carpenter and J T Chalker},
	date = {1989/07/30},
	doi = {10.1088/0953-8984/1/30/004},
	isbn = {0953-8984},
    journal = {J. Phys. Condens. Matter},
	number = {30},
	pages = {4907},
	title = {The phase diagram of a generalised {XY} model},
	url = {https://dx.doi.org/10.1088/0953-8984/1/30/004},
	volume = {1},
	year = {1989}
}

@article{Jauregui2019,
author = {Luis A. Jauregui  and Andrew Y. Joe  and Kateryna Pistunova  and Dominik S. Wild  and Alexander A. High  and You Zhou  and Giovanni Scuri  and Kristiaan De Greve  and Andrey Sushko  and Che-Hang Yu  and Takashi Taniguchi  and Kenji Watanabe  and Daniel J. Needleman  and Mikhail D. Lukin  and Hongkun Park  and Philip Kim },
title = {Electrical control of interlayer exciton dynamics in atomically thin heterostructures},
journal = {Science},
volume = {366},
number = {6467},
pages = {870-875},
year = {2019},
doi = {10.1126/science.aaw4194}
}

@article{Rivera2015,
author = {Rivera, P. and Schaibley, J. R. and Jones, A. M. and Gutiérrez, H. R. and Yan, J. and Mandrus, D. and Wang, F. and Yao, W. and Xu, X. and Kim, M. J. and Park, J. and Liebig, D. and Xu, Y. and Plumb, N. C. and Simmers, D. P. and Wallace, R. M.},
title = {Observation of long-lived interlayer excitons in monolayer {MoSe$_2$–WSe$_2$} heterostructures},
journal = {Nat. Commun.},
volume = {6},
pages = {6242},
year = {2015},
doi = {10.1038/ncomms7242}
}

@article{Rivera2016,
author = {Pasqual Rivera  and Kyle L. Seyler  and Hongyi Yu  and John R. Schaibley  and Jiaqiang Yan  and David G. Mandrus  and Wang Yao  and Xiaodong Xu },
title = {Valley-polarized exciton dynamics in a {2D} semiconductor heterostructure},
journal = {Science},
volume = {351},
number = {6274},
pages = {688-691},
year = {2016},
doi = {10.1126/science.aac7820}
}

@article{Zhang2019,
  title = {Highly valley-polarized singlet and triplet interlayer excitons in van der {W}aals heterostructure},
  author = {Zhang, Long and Gogna, Rahul and Burg, G. William and Horng, Jason and Paik, Eunice and Chou, Yu-Hsun and Kim, Kyounghwan and Tutuc, Emanuel and Deng, Hui},
  journal = {Phys. Rev. B},
  volume = {100},
  issue = {4},
  pages = {041402},
  numpages = {6},
  year = {2019},
  month = {Jul},
  publisher = {American Physical Society},
  doi = {10.1103/PhysRevB.100.041402}
}

@article{Liu2021,
author = {E. Liu and E. Barre and J. van Baren and D. S. Komatsu and D. K. Singh and S. A. Lee and Y. J. Yuan and K. Watanabe and T. Taniguchi and P. Jarillo-Herrero and D. Lu and D. A. Reis},
title = {Signatures of moiré trions in {WSe$_2$/MoSe$_2$} heterobilayers},
journal = {Nature},
volume = {594},
pages = {46–50},
year = {2021},
doi = {10.1038/s41586-021-03583-y}
}

@article{Brotons2021,
title = {Moir\'e-Trapped Interlayer Trions in a Charge-Tunable {${\mathrm{WSe}}_{2}/{\mathrm{MoSe}}_{2}$} Heterobilayer},
author = {Brotons-Gisbert, Mauro and Baek, Hyeonjun and Campbell, Aidan and Watanabe, Kenji and Taniguchi, Takashi and Gerardot, Brian D.},
journal = {Phys. Rev. X},
volume = {11},
issue = {3},
pages = {031033},
numpages = {12},
year = {2021},
month = {Aug},
publisher = {American Physical Society},
doi = {10.1103/PhysRevX.11.031033}
}

@dataset{zenodo,
  author       = {Belgaonkar, Pradyumna P. and
                  Zimmerman, Michal and
                  Gazit, Snir and
                  Orgad, Dror},
  title        = {Trimer superfluidity of antiparallel dipolar
                   excitons in a bilayer heterostructure
                  },
  month        = jul,
  year         = 2025,
  publisher    = {Zenodo},
  doi          = {10.5281/zenodo.16165694},
  url          = {https://doi.org/10.5281/zenodo.16165694},
}

@article{Macia,
  title = {Single-particle versus pair superfluidity in a bilayer system of dipolar bosons},
  author = {Macia, A. and Astrakharchik, G. E. and Mazzanti, F. and Giorgini, S. and Boronat, J.},
  journal = {Phys. Rev. A},
  volume = {90},
  issue = {4},
  pages = {043623},
  numpages = {5},
  year = {2014},
  month = {Oct},
  publisher = {American Physical Society},
  doi = {10.1103/PhysRevA.90.043623},
  url = {https://link.aps.org/doi/10.1103/PhysRevA.90.043623}
}

@article{Babaev,
  title = {Superfluid drag in the two-component {B}ose-{H}ubbard model},
  author = {Sellin, Karl and Babaev, Egor},
  journal = {Phys. Rev. B},
  volume = {97},
  issue = {9},
  pages = {094517},
  numpages = {9},
  year = {2018},
  month = {Mar},
  publisher = {American Physical Society},
  doi = {10.1103/PhysRevB.97.094517},
  url = {https://link.aps.org/doi/10.1103/PhysRevB.97.094517}
}

@article{Peeters,
  title = {Interlayer excitons in transition metal dichalcogenide heterostructures},
  author = {Van der Donck, M. and Peeters, F. M.},
  journal = {Phys. Rev. B},
  volume = {98},
  issue = {11},
  pages = {115104},
  numpages = {10},
  year = {2018},
  month = {Sep},
  publisher = {American Physical Society},
  doi = {10.1103/PhysRevB.98.115104},
  url = {https://link.aps.org/doi/10.1103/PhysRevB.98.115104}
}

@article{Calman2018,
  author    = {Calman, E. V. and Fogler, M. M. and Butov, L. V. and Hu, S. and Mishchenko, A. and Geim, A. K.},
  title     = {Indirect excitons in van der {Waals} heterostructures at room temperature},
  journal   = {Nat. Commun.},
  volume    = {9},
  number    = {1},
  pages     = {1895},
  year      = {2018},
  month     = {May},
  doi       = {10.1038/s41467-018-04293-7},
  url       = {https://doi.org/10.1038/s41467-018-04293-7}
}

@article{hBN-impurities,
  title = {Native point defects and impurities in hexagonal boron nitride},
  author = {Weston, L. and Wickramaratne, D. and Mackoit, M. and Alkauskas, A. and Van de Walle, C. G.},
  journal = {Phys. Rev. B},
  volume = {97},
  issue = {21},
  pages = {214104},
  numpages = {13},
  year = {2018},
  month = {Jun},
  publisher = {American Physical Society},
  doi = {10.1103/PhysRevB.97.214104},
  url = {https://link.aps.org/doi/10.1103/PhysRevB.97.214104}
}

@article{Seo2024,
  author    = {Seo, Yuta and Tsuji, Yuki and Onodera, Momoko and Moriya, Rai and Zhang, Yijin and Watanabe, Kenji and Taniguchi, Takashi and Machida, Tomoki},
  title     = {Spectrum of Tunneling Transport through Phonon-Coupled Defect States in a Carbon-Doped Hexagonal Boron Nitride Barrier},
  journal   = {Nano Lett.},
  volume    = {24},
  number    = {43},
  pages     = {13733--13740},
  year      = {2024},
  month     = {October},
  publisher = {American Chemical Society},
  doi       = {10.1021/acs.nanolett.4c03847},
  url       = {https://doi.org/10.1021/acs.nanolett.4c03847}
}

@article{electric-tune,
  author    = {Kistner-Morris, Jed and Shi, Ao and Liu, Erfu and Arp, Trevor and Farahmand, Farima and Taniguchi, Takashi and Watanabe, Kenji and Aji, Vivek and Lui, Chun Hung and Gabor, Nathaniel},
  title     = {Electric-field tunable {Type-I} to {Type-II} band alignment transition in {MoSe$_2$/WS$_2$} heterobilayers},
  journal   = {Nat. Commun.},
  volume    = {15},
  number    = {1},
  pages     = {4075},
  year      = {2024},
  month     = {May},
  doi       = {10.1038/s41467-024-48321-1},
  url       = {https://doi.org/10.1038/s41467-024-48321-1}
}

@article{exciton-superfluidity,
author = {Jacob Cutshall  and Fateme Mahdikhany  and Anna Roche  and Daniel N. Shanks  and Michael R. Koehler  and David G. Mandrus  and Takashi Taniguchi  and Kenji Watanabe  and Qizhong Zhu  and Brian J. LeRoy  and John R. Schaibley },
title = {Imaging interlayer exciton superfluidity in a {2D} semiconductor heterostructure},
journal = {Sci. Adv.},
volume = {11},
number = {1},
pages = {eadr1772},
year = {2025},
doi = {10.1126/sciadv.adr1772},
URL = {https://www.science.org/doi/abs/10.1126/sciadv.adr1772}
}

@article{Wang-review,
  title = {Colloquium: {E}xcitons in atomically thin transition metal dichalcogenides},
  author = {Wang, Gang and Chernikov, Alexey and Glazov, Mikhail M. and Heinz, Tony F. and Marie, Xavier and Amand, Thierry and Urbaszek, Bernhard},
  journal = {Rev. Mod. Phys.},
  volume = {90},
  issue = {2},
  pages = {021001},
  numpages = {25},
  year = {2018},
  month = {Apr},
  publisher = {American Physical Society},
  doi = {10.1103/RevModPhys.90.021001},
  url = {https://link.aps.org/doi/10.1103/RevModPhys.90.021001}
}

@article{Shanks2021,
  author    = {Shanks, Daniel N. and Mahdikhanysarvejahany, Fateme and Muccianti, Christine and Alfrey, Adam and Koehler, Michael R. and Mandrus, David G. and Taniguchi, Takashi and Watanabe, Kenji and Yu, Hongyi and LeRoy, Brian J. and Schaibley, John R.},
  title     = {Nanoscale Trapping of Interlayer Excitons in a a {2D} Semiconductor Heterostructure},
  journal   = {Nano Lett.},
  volume    = {21},
  number    = {13},
  pages     = {5641--5647},
  year      = {2021},
  month     = {July},
  publisher = {American Chemical Society},
  doi       = {10.1021/acs.nanolett.1c01215},
  url       = {https://doi.org/10.1021/acs.nanolett.1c01215},
  issn      = {1530-6984}
}

@article{Unuchek2018,
  author    = {Unuchek, Dmitrii and Ciarrocchi, Alberto and Avsar, Ahmet and Watanabe, Kenji and Taniguchi, Takashi and Kis, Andras},
  title     = {Room-temperature electrical control of exciton flux in a van der {Waals} heterostructure},
  journal   = {Nature},
  volume    = {560},
  number    = {7718},
  pages     = {340--344},
  year      = {2018},
  month     = {August},
  doi       = {10.1038/s41586-018-0357-y},
  url       = {https://doi.org/10.1038/s41586-018-0357-y}
}
\fi
\ifSM

\end{document}